# Correlation of Blocking and Néel Temperatures in Ultra-thin Metallic Antiferromagnets

Kutay Akın[a*], Hasan Piskin[a*], Ege Selvi[a], Emre Demircanlı[a,b], Şevval Arı[a,b], Mohammad Hassan Ramezan zadeh[a], Bayram Kocaman[a,c], Ozhan Ozatay[a]

[a] Bogazici University, Physics Department, Istanbul, Turkey
[b] Istanbul Technical University, Physics Engineering Department, Istanbul, Turkey
[c] Siirt University, Electrical and Electronics Engineering Department, Siirt, Turkey

*Corresponding authors: kutay.akin@boun.edu.tr, hasan.piskin@boun.edu.tr

Nonvolatile spintronics-based devices that utilize electron spin both to store and transport information face a great challenge when scaled to nano dimensions due to loss of thermal stability and stray field induced disturbance in closely packed magnetic bits. The potential replacement of ferromagnetic materials with antiferromagnets may overcome some of these issues owing to the superior robustness of sublattice spin orientations to magnetic field disturbance as long as they are kept well below the Néel temperature, which is hard to measure with conventional methods, especially in the ultrathin limit. In this work, we have employed spin pumping from a soft ferromagnetic NiFe layer into widely used ultrathin metallic antiferromagnet $Ir_{20}Mn_{80}$, FeMn, PtMn, PdMn or NiMn with thicknesses in the 0.7-3 nm range, as a probe to detect damping enhancement during magnetic phase transitions. Independent measurements of the blocking temperature with magnetometry reveal that temperature dependent shifts in the resonance peaks can also be used to measure the blocking temperature, allowing the analysis of the correlation between the Néel and blocking temperatures in trilayers with permalloy and antiferromagnetic layer separated by a 3 nm thick spacer layer. The thickness dependent characterization of thermal stability in antiferromagnets provides a key element for scalable and ultrafast antiferromagnetic spintronics.

There is a renewed interest in antiferromagnetic thin film materials with intensive research efforts to uncover their potential for memory and ultrafast data processing applications [1-3]. This is primarily due to their key advantages such as robustness against external magnetic field perturbations [4] and radiation hardness for space applications [5], terahertz range dynamics for ultrafast communication and data processing [6], nonvolatility with large magnetic anisotropy [7], possible electrical manipulation/read out [8], insensitivity to stray fields [9] and abundance in nature for low cost, low power, closely packed and complementary metal oxide (CMOS) compatible memory applications [10]. However, when their thicknesses are reduced below a certain threshold, the surface effects may dominate the magnetic characteristics [11]. Incomplete spin sublattice formation, weak antiferromagnetic exchange coupling, premature phase transitions with heating, changes in magnetic resonance conditions due to increased damping, increased electrical resistivity causing Joule heating and high-power consumption during electrical manipulation, increased inhomogeneity in the anisotropy distribution due to surface roughness and chemical intermixing at the interfaces with neighbouring layers are detrimental in the scaling of antiferromagnetic devices creating a technology roadblock [5,12,13]. Therefore, it becomes exceedingly more important to clearly identify the shortcomings brought about in the ultrathin limit and to mitigate these issues by the proper choice of materials, thicknesses, and device geometries.

Metallic antiferromagnets are expected to play a special role in the development of spintronic device technologies since they have relatively low electrical resistivity and high thermal conductivity allowing to fundamentally explore the complex interaction among charge currents, spin currents, lattice vibrations and optical excitations with circularly polarized light [14]. Conventionally, these materials have been used to magnetically harden the neighbouring ferromagnetic layers through exchange bias in spin-valve and tunnel junction devices acting as readers in magnetic hard drives [15,16]. Recently, they have been found to act as efficient spin current detectors in spin pumping experiments in addition to being prone to spin orbit torques in the ultrathin limit [17-20]. The efficiency of spin injection is quantifiable by the interfacial spin-mixing conductance, whereas the characteristic length scale over which the spin torques are effective is the spin relaxation length, which is determined by the strength of the spin-orbit coupling [21]. In addition to the ability to manipulate spin sublattice orientations by means of spin angular momentum transfer, their magnetic state is electrically detectible via tunnelling anisotropic magnetoresistance (TAMR) [22]. Since all these unique features of metallic antiferromagnets are highly dependent on their surface spin structures and the nature of their interfaces with the media for spin propagation, it is clearly desirable to systematically investigate the thickness dependence of their magnetic characteristics, especially their thermal stability for miniaturized device applications.

The simplest experimental approach to estimate the order-disorder transition temperature in antiferromagnets has been through direct magnetometry measurements or spin dependent transport measurements in bilayers with soft ferromagnets that were initially cooled in the presence of an external magnetic field to establish exchange bias [15]. By monitoring the exchange anisotropy induced hysteresis loop shifts as a function of temperature, one can indeed measure a critical point called the blocking temperature, where the loop shift is reduced to zero. This threshold corresponds to the temperature at which the interfacial magnetic moments are no longer strictly tied to the bulk moments due to their rotatable anisotropy. Although the blocking phenomenon signals a weakness in the exchange stiffness at the interface, it does not provide a reliable measure of the intrinsic loss of exchange stiffness between the antiferromagnetic moments, which occurs at the Néel temperature [23]. Unfortunately, it is quite challenging to measure the phase transition point in antiferromagnets due to their insensitivity to magnetic field application. Most available techniques such as neutron diffraction [24,25], electrical resistivity [26,27], magnetic susceptibility [28,29], Spin Seebeck effect [30] measurements and nanocalorimetry [31] are not compatible with ultrathin antiferromagnetic films.

A straightforward and rapid measurement technique has been demonstrated by Frangou and co-workers to be the detection of enhanced spin pumping at the phase transition temperature, where the antiferromagnetic layer acts as a spin sink [12]. Spin pumping is the reciprocal effect of spin transfer torque whereby a spin current is generated by magnetization dynamics and pumped into an adjacent normal metal, which can be detected via the inverse spin Hall effect. After the initial theoretical framework was available for spin pumping phenomenon, several experiments were carried out where the spin sink was a ferromagnetic layer or an antiferromagnetic layer [32-34]. If the magnetization is driven into dynamics by a variable frequency source as in a vector network analyser (VNA) at a constant external magnetic field, the resulting ferromagnetic resonance (VNA-FMR) signal obtained from $S_{21}$ transmission parameter can be used to understand the nature of the spin sink. VNA-FMR technique is well established and has been shown to give results similar to conventional FMR and pulsed inductive microwave magnetometry [35,36]. Since the magnetization dynamics is in a way coupled to the spin sink by a damping mechanism, the VNA-FMR signal linewidth acts as an indicator of changes in the magnetic order on the spin current detector layer [12]. While the damping parameter consists of many intrinsic and extrinsic contributions such as spin orbit coupling [37,38], two magnon scattering [39] and spin memory loss [40], the spin pumping contribution is distinguished from the other effects since it is enhanced when there is a transition from strongly exchange coupled spins to weakly coupled fluctuating spins at the spin sink layer.

In this work, we have performed VNA-FMR measurements to probe the spin pumping efficiency enhancement and shifts in absorption peaks at the phase transition temperature of the metallic antiferromagnet spin sink in order to determine the thickness dependence of the Néel temperature and blocking temperature in Permalloy/Cu/Antiferromagnet (AFM) trilayers. All thin films were deposited on <100> oriented Si wafers with 500 nm thermal oxide to avoid signal losses by absorption of the substrate. The multilayer stacks consisted of Ti (5) / $Ni_{81}Fe_{19}$ (10) / Cu (3) / AFM (t) / Ti (5) (where AFM layer is either PtMn, PdMn, NiMn, FeMn or $Ir_{20}Mn_{80}$ with thicknesses in between 0.7 to 3 nm. All deposition rates were calibrated with surface profilometry and thicknesses are given in nm.) Ti was used both as an adhesion layer and oxidation passivation cap layer whose oxide is also known to have low spin absorption characteristics [41]. The multilayers were deposited in a dc magnetron sputter system with a base pressure lower than $5 *10^{-8}$ Torr. During sputtering 2mTorr pressure was maintained at 5sccm Ar flow rate. The substrates were rotated at 30 rpm for thickness uniformity and exposed to 250 Oe magnetic field to define an anisotropy axis parallel to the signal line of the measurement cell.

Fig. 1(a) shows a schematic drawing of the VNA-FMR measurement setup. A home-made radio frequency (RF) insert made of semirigid coaxial cables is placed inside a cryogenic chamber capable of applying magnetic fields up to 9 T along the y axis (Quantum Design Dynacool Physical Property Measurement System, PPMS). A home-made measurement cell consists of a 50 Ω coplanar waveguide (CPW) soldered on a brass sample holder that provides standard Ground-Signal-Ground (GSG) RF connections to Sub Miniature Version A (SMA) connectors. The device under test is a continuous thin film mounted on the CPW upside down with a Kapton tape (the so-called flip-chip method). While the magnet inside the chamber provides a dc magnetic field along the y-axis, the VNA (Keysight P9374A) connected to the RF insert as shown in Fig. 1(a) acts as a swept frequency RF generator so that the RF current passing through the signal line produces an

RF Oersted Field along the x axis and perpendicular to the dc field. The RF absorption spectra are recorded by monitoring the forward transmission signal $S_{21}$ at dc fields from 100 Oe to 400 Oe with 25 Oe steps in the temperature range 10 K – 300 K with 10 K steps. At each stable temperature, the sample is initialized by the application of 600 Oe magnetic field in +y direction to saturate the magnetization. Then the field is reduced to zero. The source frequency is swept between 0 to 7 GHz with 3kHz steps. The applied RF power is low enough (15 dBm) to ensure no local heating, which together with the thermal reservoirs provided by the brass holder implies no local heat gradients and therefore no spin Seebeck effect contribution to the measurements. Fig. 1(b) shows the absorption spectrum for Ti (5) / $Ni_{81}Fe_{19}$ (10) / Cu (3) / Ti (5) (in nm) sample measured at 10 K in an external field of 200 Oe together with a Lorentzian fit used to quantify the resonance peak frequency and resonance linewidth (full-width half maximum, FWHM). The inset shows the shift in the resonance frequency with increasing applied field. By fitting the resonance frequency vs. magnetic field data to Kittel equation [42] (see Fig. S1) we obtain an effective magnetization of $4\pi M_{eff} = 9$ kG for Landé factor $g = 2.0$.

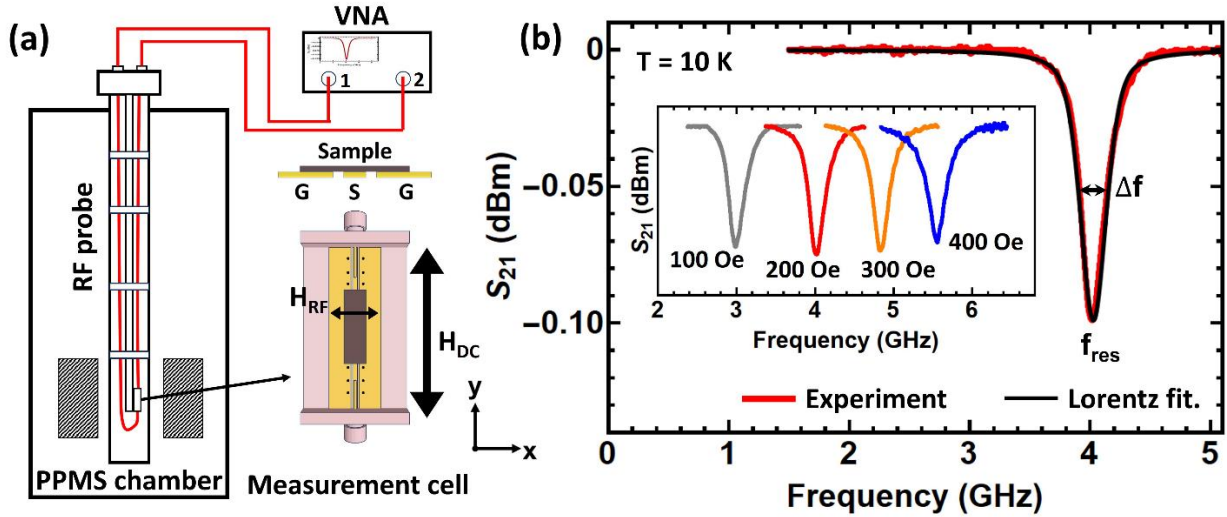

FIG. 1. (a) Cryogenic VNA-FMR measurement setup. (b) Absorption spectrum of Ti (5) / NiFe (10) / Cu (3) / Ti (5) (in nm) sample measured at 10 K in an external field of 200 Oe applied parallel to the signal line (solid red curve), the solid black curve is the Lorentzian fit used to determine FWHM (Δf). Inset: Absorption spectra measured at 10 K with an applied field of 100 Oe (gray solid curve), 200 Oe (red solid curve), 300 Oe (orange solid curve), 400 Oe (blue solid curve).

In-plane magnetic hysteresis loops with external field applied along the y-axis (the easy axis defined during deposition) were obtained using a variable temperature vibrating sample magnetometer (VSM). Measurements in the temperature range of 10 – 400 K after zero field cooling to 10 K reveal the presence of a weak exchange bias despite the 3nm thick Cu spacer layer in all sample types. Depending on the nature of the non-magnetic metal spacer, exchange bias has been shown to be a long-range interaction between the antiferromagnetic and ferromagnetic interfacial spins mediated dominantly by indirect exchange - RKKY interaction [43,44] which is more pronounced in the low temperature regime [45,46]. Fig. 2(a) shows hysteresis loops for Ti (5) / Ni$_{81}$Fe$_{19}$ (10) / Cu (3) / Ir$_{20}$Mn$_{80}$ (1) / Ti (5) (nm) multilayer sample measured at 10 K (blue) and 50 K (red). The exchange bias field $H_{ex}$ is defined to be the shift of the symmetry axis of the hysteresis loop (20 Oe at 10 K). By plotting $H_{ex}$ as a function of temperature with 10 K steps (see Fig. 2(a) inset) a critical temperature where the exchange bias drops to zero (the blocking temperature $T_B$) is found to be 50 K. By performing this sequence of measurements for samples with the same film structure but different IrMn thicknesses in 0.7 to 3 nm range, we observe a power law dependence of the blocking temperature on IrMn thickness (see Fig. 2(b)), where the gradual increase in the blocking temperature with increasing IrMn thickness is well fit to the thermally fluctuating spins model [47] except for 0.7 nm IrMn thickness. In this model, the AFM thickness ($t_{AFM}$) dependence of the blocking temperature is given as:

$$\frac{T_B(\infty) - T_B(t_{AFM})}{T_B(\infty)} = \left(\frac{\xi_0}{t_{AFM}}\right)^\delta, \tag{1}$$

where $T_B$ ($\infty$) indicates the blocking temperature of the bulk AFM, the shift exponent $\delta = 1/2\lambda$ with $\lambda$ being the critical index of the temperature-dependent AFM moment $m_{AFM}(T) = m_{AFM}^{T=0}(1 - T/T_N)^\lambda$, where $T_N$ is the Néel temperature [48]. $\xi_0 = J_{int}/2K_{AFM}ra$ corresponds to the correlation length, where $J_{int}$ is the interfacial exchange coupling between the FM and the AFM spins, $K_{AFM}$ is the magnetic anisotropy constant, and $r$ and $a$ are the grain size and the lattice constant of AFM respectively [23]. From the fit, we have found the shift exponent to be 0.63 ± 0.01 and the correlation length to be 0.87 ± 0.01 nm. $T_B$ ($\infty$) is assumed to be 560 K [49]. The discrepancy between the data and the model, for lowest IrMn thickness can be attributed to incomplete spin sub-lattice formation at 0.7 nm IrMn thickness, indicating that the antiferromagnetic exchange coupling between neighbouring moments is not yet well established. Indeed, polycrystalline

Ir$_{20}$Mn$_{80}$ is known to crystallize in a cubic structure with a lattice constant a =3.78 Å [50] and (111) texturing implying that 0.7 nm film thickness corresponds to less than 2 monolayers of IrMn and furthermore, this thickness is below the correlation length obtained from thermal fluctuating spins model which is 0.87 nm.

VNA-FMR measurements (repeated in the temperature range 10 to 400 K) on Ti (5) / Ni$_{81}$Fe$_{19}$ (10) / Cu (3) / Ir$_{20}$Mn$_{80}$ (1) / Ti (5) (nm) multilayer (the same structure as in VSM measurements presented in Fig. 2(a)) show a divergence of the resonant absorption peaks in the presence of +100 Oe and -100 Oe external fields applied along the y-axis for temperatures below 50 K [Fig. 2(c)]. This raises the question whether the asymmetric behaviour up to a critical temperature can be interpreted as an alternative means of measuring the blocking temperature. The blocking temperature values detected by using the VNA-FMR technique appear to be in good agreement [Fig. 2(d)] with those obtained from hysteresis loops [Fig. 2(b)] reconfirming the fit (with fit parameters of correlation length: 0.90 ± 0.03 nm and shift exponent: 0.69 ± 0.04) to the thermally fluctuating spins model and also reconfirming the discrepancy at the lowest IrMn thickness. Therefore, indeed the asymmetric temperature dependent behaviour of VNA-FMR resonant peaks with change in polarity of the external field is proven to be a reliable method of determining the blocking temperatures in samples with ultra-thin antiferromagnetic layers.

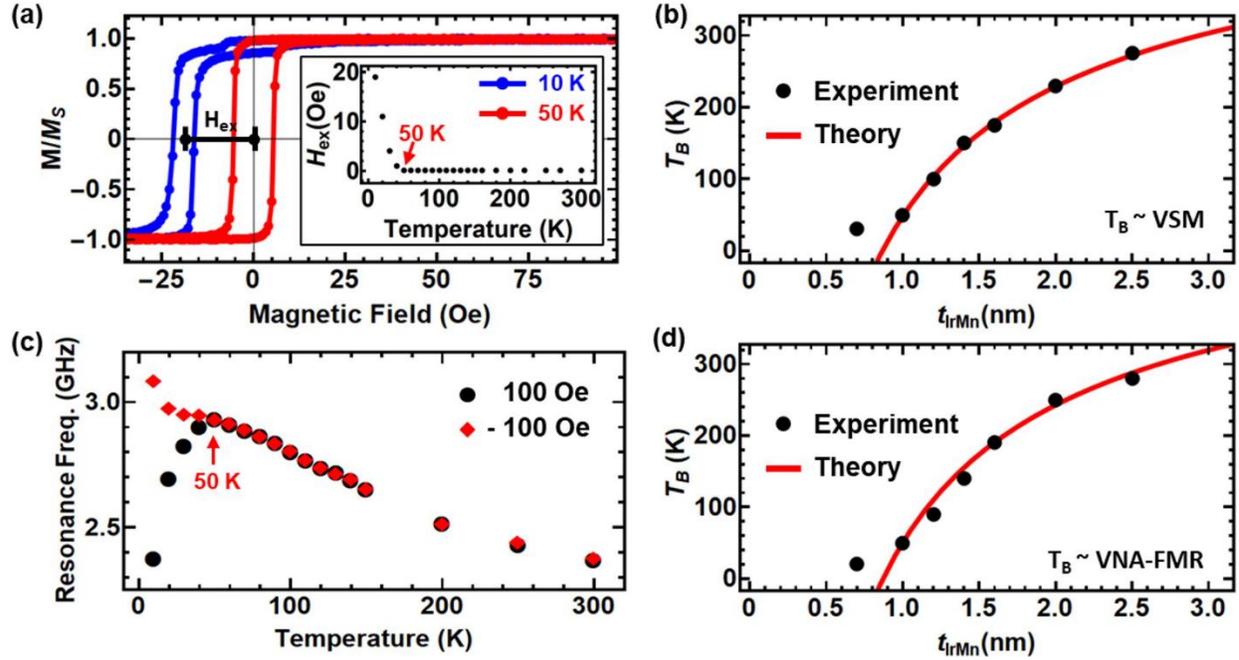

FIG. 2. (a) VSM hysteresis loops for Ti (5) / Ni$_{81}$Fe$_{19}$ (10) / Cu (3) / IrMn (1) / Ti (5) (in nm), blue at 10 K, red at 50 K. Inset: Exchange bias field vs. temperature displaying blocking at 50 K. (b) Blocking temperature vs Thickness (IrMn) (VSM hysteresis loops method), black dots, theoretical fit (Thermal Fluctuation Model) red curve. (c) VNA-FMR resonance frequency vs. temperature, dots at +100 Oe, diamonds at -100 Oe magnetic field applied along the y-axis. (d) Blocking temperature vs Thickness (IrMn) (VNA-FMR resonance frequency method), black dots, theoretical fit (Thermal Fluctuation Model) red curve.

The blocking temperatures for the Ti (5) / Ni$_{81}$Fe$_{19}$ (10) / Cu (3) / AFM (t) / Ti (5) (nm) (t ~ 1 – 3 nm) structure were determined as a function of AFM thickness for PtMn, PdMn, NiMn and FeMn spin sinks as shown in Fig. 3. Fitting the experimental data to the thermal fluctuation model yields a correlation length of $\xi = 0.52 \pm 0.04, 0.96 \pm 0.03, 0.91 \pm 0.13, 1.31 \pm 0.01$ nm and shift exponent of $\delta = 0.05 \pm 0.01, 0.14 \pm 0.01, 0.10 \pm 0.02, 1.52 \pm 0.07$ for PtMn ($T_B (\infty) = 616$ K) [51], PdMn ($T_B (\infty) = 520$ K) [52], NiMn ($T_B (\infty) = 513$ K) [53] and FeMn ($T_B (\infty) = 425$ K) [54] respectively. It is worth mentioning that these values are for multilayers with ultra-thin antiferromagnets as deposited in 250 Oe magnetic field and are prone to change with any post deposition treatments applied to the AFM layers such as annealing at a temperature above the paramagnetic phase transition point in an external magnetic field, ion irradiation and plasma treatment in addition to changes in the deposition conditions such as depositing to hot substrates or substrates with different crystal orientations etc. [55].

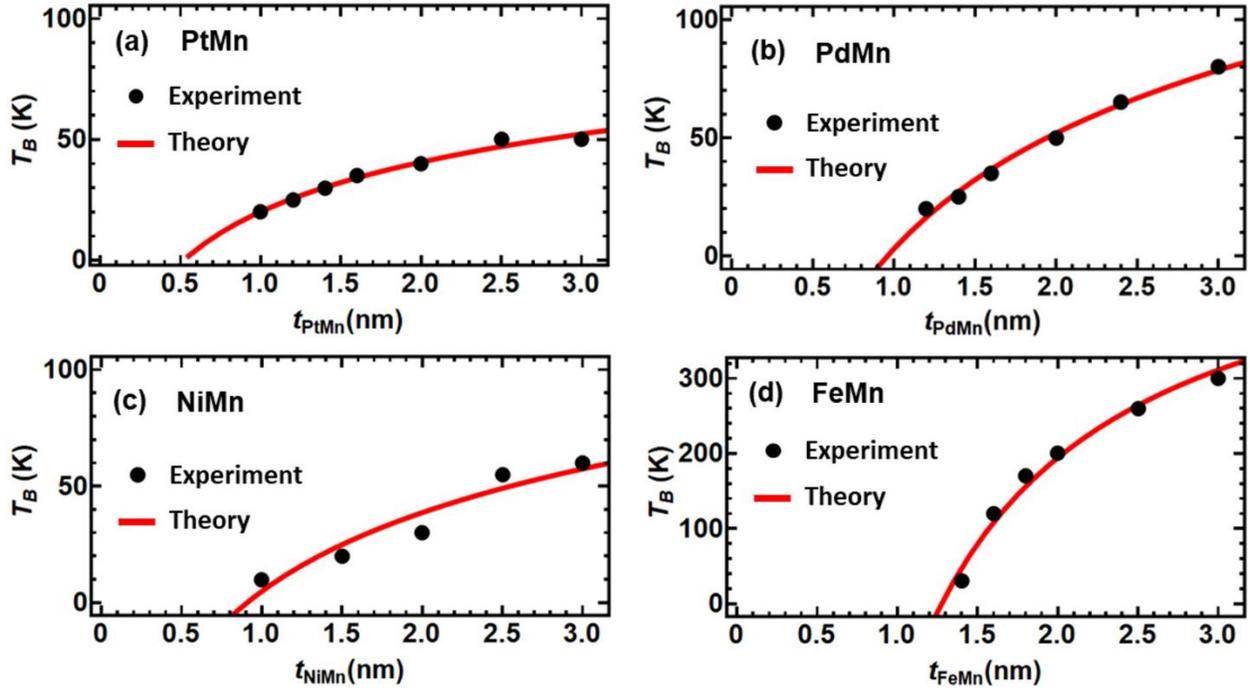

FIG. 3. Blocking temperature vs AFM Thickness (VNA-FMR resonance frequency method), black dots, theoretical fit (Thermal Fluctuation Model) red curve for the structure Ti (5) / $Ni_{81}Fe_{19}$ (10) / Cu (3) / AFM (t) / Ti (5) (nm) where the AFM is PtMn (a), PdMn (b), NiMn (c) and FeMn (d).

The temperature dependence of resonant absorption of microwave power as a result of spin pumping into the antiferromagnetic spin sink, not only allows deduction of the blocking temperature $T_B$ from the resonance frequency peak but also the Néel temperature $T_N$ from linewidth $\Delta f$ measurements. This has been shown to be due to a significant increase in the Gilbert damping parameter of the multilayer structure during a phase transition [12]. To achieve independent measurements of the blocking temperature for interfaces with NiFe /Cu /AFM structure and Nèel temperatures for ultrathin metallic antiferromagnetic layers of interest in this work, we have built multilayer structures composed of Ti (5) / $Ni_{81}Fe_{19}$ (10) / AFM (t) / Cu (3) / Ti (5) (in nm) where t is in the range 0.7 to 3 nm. First, to verify the validity of our measurement and data analysis protocol in comparison to previous work of Frangou et al., we have used $Ir_{20}Mn_{80}$ as the AFM layer. The thickness dependence of the Néel temperature for ultrathin IrMn obtained from temperature dependent VNA-FMR linewidth measurements for t = 0.7, 1.0 and 1.4 nm is shown in comparison to the results of Frangou et al. [12] in the supplementary figure [Fig. S2]. Here, the apparent Gilbert damping parameter $\alpha$ is obtained using Eq. (2) [56]

$$\alpha = \frac{2\pi \Delta f}{\gamma \mu_0 \left[2\left(H_{ext} + H_{uni}\right) + M_{eff}\right]}, \qquad (2)$$

where $\Delta f$, $H_{ext}$, $H_{uni}$, $M_{eff}$, $\gamma$ are the full-width half maximum of the Lorentzian fit, external applied field, uniaxial anisotropy field, effective magnetization, and gyromagnetic ratio respectively. In this experiment the external magnetic field is kept constant at 200 Oe, and $M_{eff}$ is recalculated using the fits to the Kittel Equation [Eq. (S1)] at each temperature. For a continuous permalloy thin film the uniaxial anisotropy field is negligibly small. The apparent damping has several intrinsic and/or extrinsic contributions resulting from interband and intraband transitions in permalloy in addition to interfacial strong spin orbit coupling induced by the AFM layer during spin pumping [57]. The spin pumping contribution $\alpha^p$ can be extracted by subtracting $\alpha$ values obtained from samples with no AFM layer (t = 0 nm) from the data with non-zero AFM layer thickness as shown in the inset of Fig. 4(a) for PtMn. This procedure was repeated by altering the AFM thickness for different antiferromagnetic materials namely, PtMn, PdMn, NiMn, and FeMn, as illustrated on the top windows of Fig. 4(a), (b), (c), and (d) respectively. Dotted red lines are added to guide the eyes and demonstrate the background on each $\alpha^p$. After subtracting the background, the temperature dependent enhancement of spin pumping-originated damping values ($\delta \alpha^p$) are obtained as shown at the bottom windows of Fig. 4(a), (b), (c), and (d) for PtMn, PdMn, NiMn and FeMn respectively.

The thermally induced relative change in the damping values can then be used to calculate the spin mixing conductance for each Cu/AFM interface as a measure of the spin transparency of these interfaces using Eq. (3)

$$\frac{\delta g_{eff}^{\uparrow \downarrow}}{S} = \frac{4\pi M_S\, t_{NiFe}}{|\gamma|\hbar} \delta \alpha^p, \qquad (3)$$

where $(\delta g_{eff}^{\uparrow \downarrow}/S)$ is the effective spin mixing conductance divided by the norm of the spin operator and $M_S$, $t_{NiFe}$, $\hbar$, $\gamma$ denote the saturation magnetization of the ferromagnetic layer, the thickness of the ferromagnetic layer, reduced Planck constant, and gyromagnetic ratio respectively. The saturation magnetization of our NiFe layer is taken as 600 kAm$^{-1}$. For our specific trilayer structure, it is shown in [58] that effective spin mixing conductance is approximately equal to the spin mixing conductance of the Cu/AFM interface ($g_{eff}^{\uparrow\downarrow} \sim g_{Cu/AFM}^{\uparrow\downarrow}$). The temperature dependent variation of the spin mixing conductance (the right axes) is shown at the bottom windows of Fig. 4 (a), (b), (c), and (d) for Cu/PtMn, Cu/PdMn, Cu/NiMn and Cu/FeMn interfaces respectively.

The critical temperature (Néel temperature) at which the maximum damping enhancement occurs is obtained from $\delta\alpha^p$ vs Temperature plots of Fig. 4. Fig. 5 shows the dependence of the Néel temperature on the AFM layer thickness for PtMn, PdMn, NiMn, and FeMn in 5 (a), (b), (c), and (d) respectively. It has been shown that by using the mean field approximation, critical temperature change with the material thickness can be described as a linear function for the epitaxial ultra-thin films with known interatomic distance, Bulk critical temperature, and range of spin-spin interactions [59]. Linear fits have been performed to the data shown in Fig. 5 (shown as red lines) following Eq. (4) [59]

$$T_N(t_{AFM}) = T_{N,Bulk} \frac{(\frac{t_{AFM}}{d} - 1)}{2N_0}, \qquad (4)$$

where $t_{AFM}$, $d$, and $N_0$ denote AFM thickness, interatomic distance, and spin-spin interaction range respectively. These fits suggest an average spin-spin interaction range $N_0$ (in monolayers) of 6 and 13 for PdMn and NiMn respectively. However, the values suggested by these fits are unphysical (too large (more than 10000) for PtMn and too small (less than a monolayer) for FeMn). In an effort to understand the origin of this disagreement with the mean field expression of Eq. (4) we have performed additional structural characterization of all AFM layers.

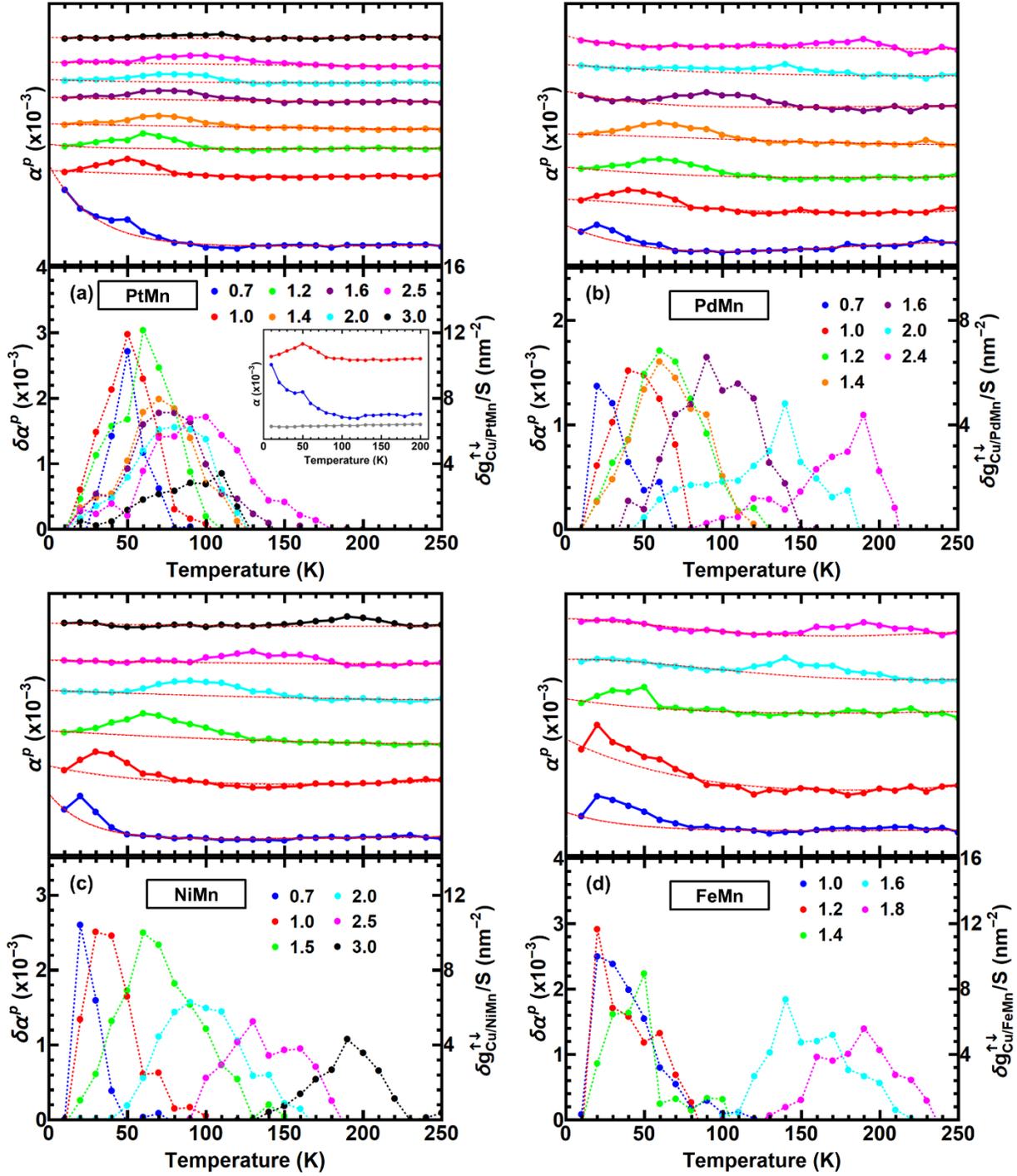

FIG. 4. Temperature dependence of spin pumping contribution to damping parameter $\alpha^p$ (top window) and its deviation $\delta\alpha^p$ from the background (shown as red dotted curves in the top windows) (bottom window - left axis) together with the thermally induced enhancement in normalised spin mixing conductance (bottom window - right axis) for different AFM thickness values (given in nm) in the structure Ti (5) / $Ni_{81}Fe_{19}$ (10) / Cu (3) / AFM (t) / Ti (5) (nm), where the AFM is PtMn (a), PdMn (b), NiMn (c), and FeMn (d). The inset in (a) illustrates the temperature dependence of apparent damping $\alpha$ values for t = 0 (gray), 0.7 (blue) and 1 (red) nm thick PtMn multilayers. All experimental data points (dots) are shown with linear connections as a guide to the eye.

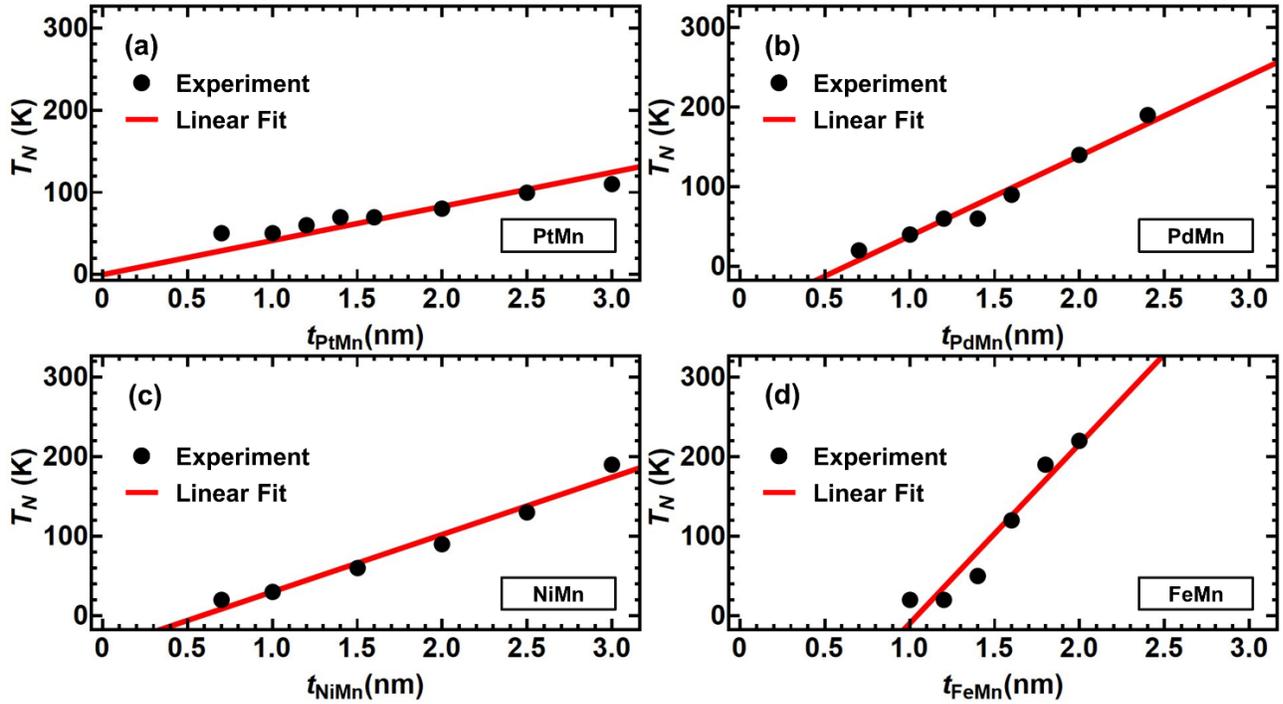

FIG. 5. Néel temperature vs AFM Thickness, experimental data - black dots, linear fit to mean field approximation – red lines for Ti (5) / Ni$_{81}$Fe$_{19}$ (10) / Cu (3) / AFM (t) / Ti (5) (nm), where the AFM layer is (a) PtMn, (b) PdMn, (c) NiMn, (d) FeMn.

The x-ray diffraction (XRD) patterns of 50 nm thick AFM materials (fabricated on a (111) oriented Pt (5 nm) buffer layer) were obtained by using a Rigaku D/Max-Ultima diffractometer ($\lambda$ = 1.5406 Å, CuK$_\alpha$). The data show that FeMn and PtMn films exhibit a polycrystalline phase with (111) and (002) orientations although they have a (111) oriented Pt (5 nm) buffer layer [60,61] [Fig. 6], while the IrMn, NiMn and PdMn films have the (111) oriented single crystal phase as shown in the supplement [62-64] [Fig. S3]. Each observed peak have been analysed by using Debye-Scherrer equation to get average crystallite sizes present throughout the thin film structures [65]:

$$D = \frac{k\lambda}{\beta cos\theta}, \quad (5)$$

where $D$ is the average crystal size (nm), $k$ is Scherrer's constant (it can take value between 0.8 and 2.08, in this study $k = 1.35$ constant), $\beta$ is the full width at half maximum of a related peak and $\theta$ is the diffraction angle. According to Eq. (5), FeMn film has average crystallites in the size of 7.82 nm for (111) orientation and 7.32 nm for (002), while PtMn film has 21.12 nm for (111) and 9.25 nm for (002) orientations. The calculated average crystallite sizes for the remaining three AFMs are presented in the supplementary file.

The XRD analysis clearly proves that FeMn and PtMn display peaks with comparable intensities for a mixture of two crystal orientations with relatively small average crystallite sizes. We attribute the discrepancy between the mean field expression and the data for FeMn and PtMn to the fact that the theory is strictly valid for epitaxial films with a single crystal orientation [59]. This however does apply to the cases of PdMn and NiMn as they display single crystalline phases with relatively large average crystallite sizes as shown in the supplementary file.

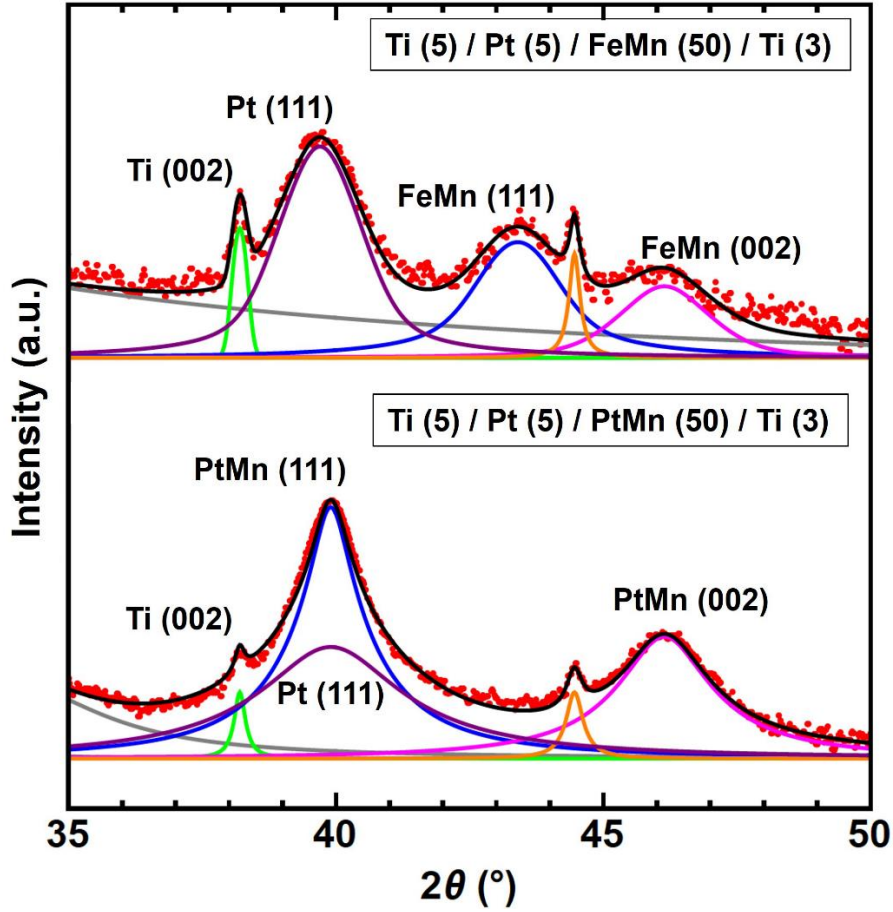

FIG. 6. XRD patterns for FeMn (top) and PtMn (bottom). All peaks are labelled except the sample holder peak ($2\theta = 44.46°$, orange) present in all measurements. The sum of all the Lorentzian fits is given by the black curve, which closely matches the experimental data given by the red dots. All the thicknesses are given in nm.

In conclusion, we have shown that in Ferromagnet ($Ni_{81}Fe_{19}$)/ Nonmagnetic spacer (Cu) /Antiferromagnetic metal (AFM) multilayer structures, provided that the spacer layer is thin enough to allow indirect exchange coupling, independent measurements of the blocking temperature and Néel temperature can be made simultaneously using the VNA-FMR spin pumping technique for ultrathin AFM spin-sink layers. While temperature dependent shifts in the resonance frequency allows determination of the blocking temperature (also verified by VSM measurements), enhanced damping obtained from thermally induced increases in resonance linewidth yields the Néel temperature. Thickness dependence of the critical temperatures $T_B$ and $T_N$ for the widely used antiferromagnetic metals $Ir_{20}Mn_{80}$, PtMn, PdMn, NiMn and FeMn allow the quantification of the correlation length (minimum thickness for establishment of exchange bias at the interface) and spin-spin correlation length (minimum thickness required for the establishment of a strong antiferromagnetic exchange coupling in the bulk AFM) when fit to thermal fluctuation model and linear mean field approximation expression respectively. However, the theoretical fits for the thickness dependence of Néel temperatures of PtMn and FeMn display a significant discrepancy with experimental data points leading to unreliable suggested correlation lengths due to the existence of two crystal orientations with comparable grain sizes as determined by XRD measurements. The enhanced spin pumping during the phase transition is also quantified by the change in the spin mixing conductance at the Cu/AFM interface. The accessibility of critical phase transition temperatures at the bulk and at interfaces with normal metals and ultrathin metallic antiferromagnets with a single VNA-FMR spin pumping measurement provides the key ingredients in identifying the role of antiferromagnets in the spintronic device technology roadmap. Unprecedented potential benefits of antiferromagnetic devices can only be realized with the proper handling of thermal stability in the ultrathin limit within the realm of scalable, disruptive antiferromagnetic device technologies.

**Acknowledgement**

This work was supported by the Scientific and Technological Research Council of Turkey (TUBITAK) under the contract no. 118F431 & Bogazici University Research Fund under the contract no. 20B03M6.

K.A. and H.P. contributed equally to this work.

**Supplementary**

**Correlation of Blocking and Néel Temperatures in Ultra-thin Metallic Antiferromagnets**

Kutay Akın[a], Hasan Piskin[a], Ege Selvi[a], Emre Demircanlı[a,b], Şevval Arı[a,b], Mohammad Hassan Ramezan zadeh[a], Bayram Kocaman[a,c], Ozhan Ozatay[a]

[a] Bogazici University, Physics Department, Istanbul, Turkey
[b] Istanbul Technical University, Physics Engineering Department, Istanbul, Turkey
[c] Siirt University, Electrical and Electronics Engineering Department, Siirt, Turkey

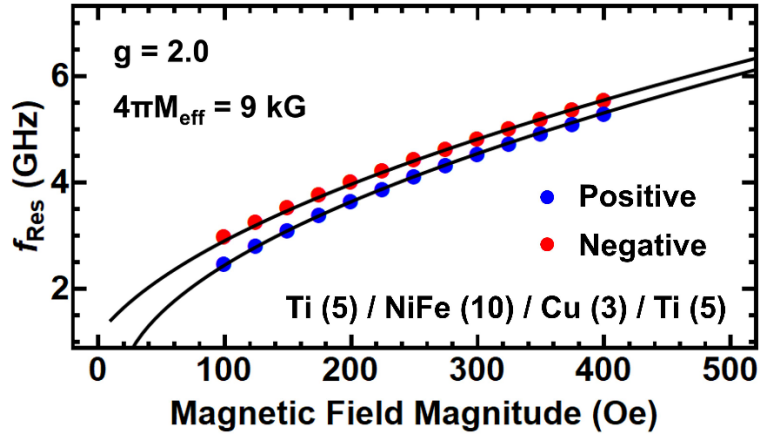

FIG. S1. Resonance frequency vs magnetic field magnitude for Ti (5) / NiFe (10) / Cu (3) / Ti (5) (nm) at 10K. Blue dots: Magnetic field in +y direction. Red dots: Magnetic field in -y direction. Solid black line: Kittel-Curve fit obtained from Eq. (S1). The deviation between positive and negative field scans is due to the remanent field of the magnet (~ 17Oe).

$$f = \frac{\gamma}{2\pi}\sqrt{H_{res}(H_{res} + 4\pi M_{eff})} \qquad (S1)$$

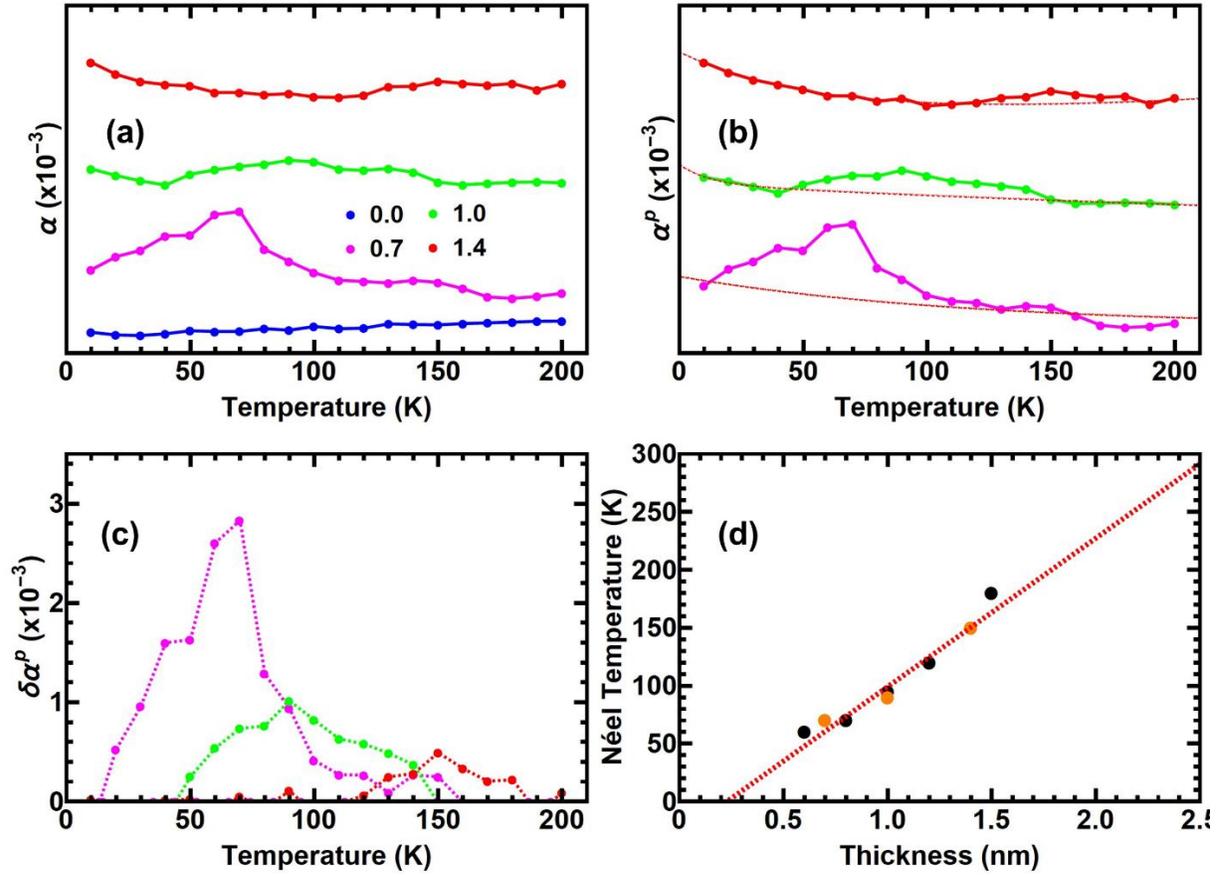

FIG. S2. (a) Apparent damping values for the Ti (5) / Ni$_{81}$Fe$_{19}$ (10) / Cu (3) / IrMn (t) / Ti (5), t= 0, 0.7, 1.0, 1.4 (in nm) samples. (b) Pumping originated damping values for the samples with non-zero AFM thickness, with their guiding dotted background values. (c) Induced variation of the damping parameter due to the phase change of the AFM layer near its critical temperature. (d) Néel temperature vs thickness graph for the Ni$_{81}$Fe$_{19}$ (10) / Cu (3) / IrMn (t) trilayer systems (black dots and the fit is taken from [12], orange dots from this work).

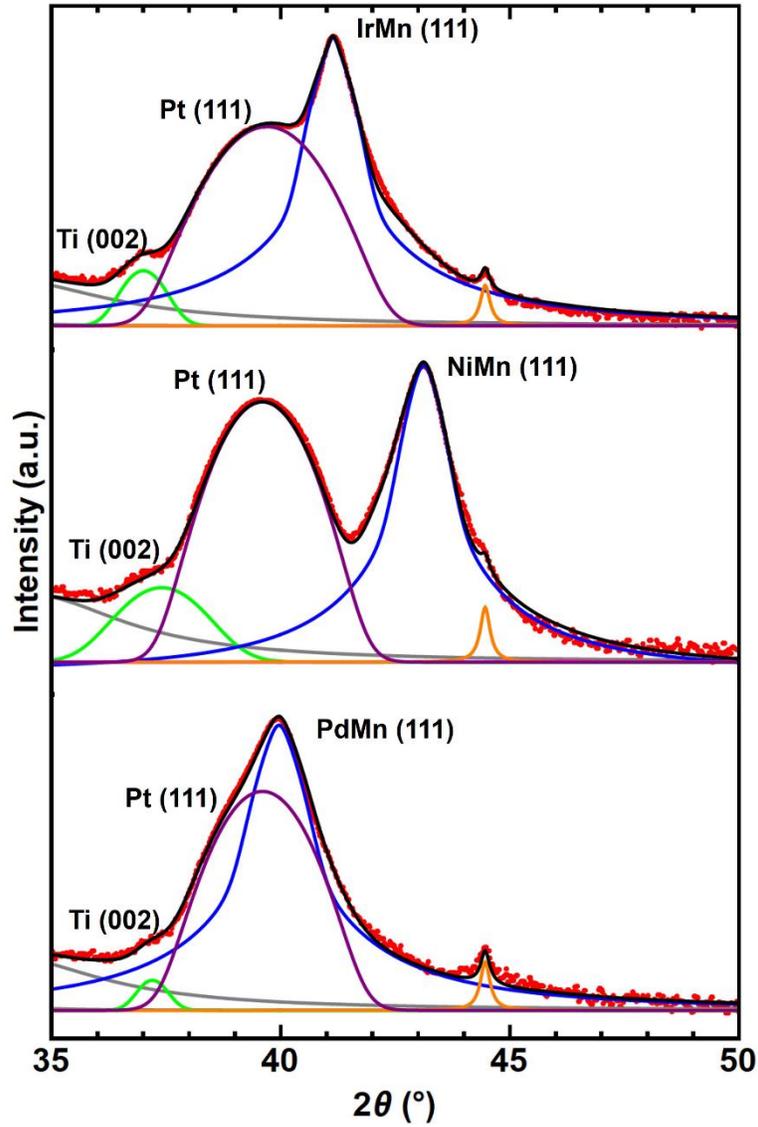

FIG. S3. XRD patterns for IrMn and PdMn, NiMn, with logarithmic y-axes. Samples are displaying single crystal structures with average crystallite sizes of 26.52, 18.65, 20.02 (in nm) correspondingly, which are calculated by the Eq. (5) of the article.